\documentclass[fleqn,10pt]{wlscirep}
\usepackage{epsfig,graphicx}

\title{Statistical Dynamics of Regional Populations and Economies}
\author[1,2]{Jie Huo}
\author[1,2,*]{Xu-Ming Wang}
\author[1,2]{Rui Hao}
\author[1]{Peng Wang}
\affil[1]{School of Physics and Electronic-Electrical Engineering, Ningxia University, Yinchuan,750021, P. R. China}
\affil[2]{Ningxia Key Laboratory of Intelligent Sensing for Desert Information, Yinchuan, 750021,P. R. China}

\affil[*]{wang$_{-}$xm@126.com(Xu-Ming Wang)}


\begin{abstract}
Quantitative analysis of human behavior and social development is becoming a hot spot of some interdisciplinary studies. A statistical analysis on the population and GDP of 150 cities in China from 1990 to 2013 is conducted. The result presents the cumulative probability distribution of the populations and that of the GDPs obeying the shifted power law, respectively. In order to understand these characteristics, a generalized Langevin equation describing variation of population is proposed based on the correlation between population and GDP as well as the random fluctuations of the related factors. The equation is transformed into the Fokker-Plank equation to express the evolution of population distribution. The general solution demonstrates a transform of the distribution from the normal Gaussian distribution to a shifted power law. It in fact suggests a critical point of time at which the transform takes place. The shifted power law distribution in the supercritical situation is qualitatively in accordance with the practical result. The distribution of the GDPs is derived from the well-known Cobb-Douglas production function. The result presents a change, in supercritical situation, from a shifted power law to the Gaussian distribution. This is a surprising result--the regional GDP distribution of our world will be the Gaussian distribution some day in the future. The discussions based on the changing trend of the economy growth suggest it will be true. Therefore, these theoretical attempts may draw a historical picture of our society in the aspects of population and economy.

\end{abstract}

\begin{document}

\flushbottom
\maketitle
 
\thispagestyle{empty}

\section*{Introduction}

In the past decade, migration behavior of population has drawn great interest of scientists. Many physicists involved in the studies to discover the characteristics and corresponding mechanism. They have explored the laws that dominate the relationship between the migrated population and the related factors, and studied the distributions of the related aspects such as migrated population, migrating distances and per economic incomes, etc. The groundbreaking field work based on practical observations was contributed by E. G. Ravenstein in 1885; and the well-known 7-rules that govern population migration has been proposed\cite{ravenstein1885laws}. 

From then, scientists have investigated this issue from two different points of view. One focuses on finding out the driving factor of the migrations based on the knowledge of economics, physiology and sociology; the other emphasizes extracting regularities of migration from observed data and revealing their dynamic origin. The former is represented by the following investigations. The aforementioned factors, associating in some way, are perceived as contributing to migration. Some combinations play the part of repelling force and some others serve as appealing force acting on the immigrating persons\cite{heberle1938causes}. The so-called push-pull theory was proposed based on the motivation of migration, improving the quality of living\cite{bogue1959internal}. The clearest description of this theory was suggested by Lee via taking the driving factors, amount of persons and directions, into account\cite{lee1966theory}.  A similar contribution was made by Schultz who regarded the migration as a decision-making behavior based on the income-output\cite{schultz1961investment}. The latter is denoted by investigations to reveal the physical essence of migrations. One typical representative is the well-known gravity model, which is in a form analogous to that of the gravity law\cite{zipf1946p}. It states that the number of people that migrate from one region to another is usually proportional to the product of the numbers of people living in this two regions, and inversely proportional to the distance between the two regions. This work is actually the field-broken one of the dynamic investigations. Following this idea, the intervening opportunities model was proposed to reveal how the number of visitors to a given distance relates to the number of opportunities at that distance and to the number of intervening opportunities\cite{stouffer1940intervening}. The probabilistic model for urban travel was developed to emphasize the influence of heterogeneity of individuals on the choice of destinations\cite{domencich1975urban}.
 
Gradually, many studies turn to focus on dynamic details of human movements. The most remarkable one is that practical investigation on the traveling traces via tracking bank notes and the corresponding theoretical interpretation, conducted by Brockman in 2006\cite{brockmann2006scaling}. It induced a new climax in exploring the spatial characteristics of human movements and the mechanics behind\cite{gonzalez2008understanding}. With the aid of the modern information technology, the researchers can precisely capture human motions. For instances, the data obtained by tracking cellphone calls and that by tacking short-message services has proved that the movements of the cellphone users can be characterized by a Truncated Power Law (TPL) distribution of the trip distances \cite{gonzalez2008understanding,han2011origin,candia2008uncovering,wei2009heavy,song2010modelling,wu2010evidence,zhi2011empirical}; the GPS data of taxis serves showed that the movements of the passengers can be described by an exponential distribution of the travel distances \cite{rambaldi2007mobility,jiang2009,bazzani2010statistical,jiang2011exploring,liang2012scaling,peng2012collective,riccardo2012towards}.

There have been many theoretical attempts to interpret the essential origins of the aforementioned regularities. As far as methodology is concerned, two typical methods are worth mentioning here. The first is dynamical simulation based on constructing the rules to mimic the mobility of human. For instances, the so-called continuous-time random walk model was suggested to reproduce the movements of the bank notes, and drew a conclusion that the motion of the bank note holder can be regarded as a L\'{e}vy flight\cite{brockmann2006scaling}; the preferential return (exploring a new place with a probability and returning to a preferentially visited location with the complementary probability) model was proposed to understand the individual mobility patterns uncovered by cell phone calls\cite{candia2008uncovering}, and the results match well the individual’s daily activity; a similar model, the regular mobility model, was developed to mimic the periodicity and regularity of human daily commuting mode, "home-work-leisure-home''\cite{xiao2011exact}, and the simulations are in good accordance with those practical observations presented in Refs.\citen{gonzalez2008understanding} and \citen{yan2013diversity}; the dynamical model for population migration, driven by economic income and hindered by the cost relating to displace and employment opportunity, can capture the main empirical features, mentioned above, such as the distributions of displaces and population migration sizes, etc\cite{Huo2016Statistical}. The second theoretical method is based on some related physical abstract or reduction of human mobility, and/or utilizing some fundamental principle of Physics to characterize human’s behavior referring to the knowledge of the corresponding physical system. The following efforts may be most typical. In order to understand statistical characteristics of population, the idea of the Maxell-Boltzmann statistics was adopted to interpret the TPL distribution of travel distances. This consideration may be due to the fact that there is similarity between the movement of persons in a city and that of particles in a thermodynamic system. The most probable distribution of equilibrium state can be obtained based on the principle of maximum entropy with the gross cost constraint and the assumptions of the nearly independence and the identity of particles\cite{tao2013statistical}. To reveal the mechanism dominating scaling properties of human and animal mobility, and to explain what causes the value difference of the scaling exponent between human and animal, the home-return l\'{e}vy flight model optimized by information entropy has been built. The exponents can be obtained via optimizing under the constraint of maximizing the information entropy, and indicate that the difference depends on whether there is a home to return, yes for human, not for animal\cite{hu2011toward}. The process that some persons leaving home-town and looking for better jobs incarnated by high income, short working time, good working condition, etc., can be regarded as a stochastic process just as the radiation of some particles that are emitting from an origin location, diffusing to look for some locations to release or transport the carried energies or masses. The radiation model catches such a similarity of the physical essence in these two processes. And it works well to predict the population migration from one location to another\cite{simini2012universal}. From the discussions above, it can be clear seen if an investigation can grasp analogies between the behaviors of individual/social system and that of particle/physical system, the essential description of human activity may be obtained by translating the corresponding theory or principle of physics. Here we also emphasize another aspect apparently existed in the aforementioned investigations, that is, the scaling distribution feature of human motions may be jointly determined by the stochastic and deterministic driving forces. In fact, similar mechanisms exist widely, in nature, to dominate evolution of systems\cite{wang2012statistical,rinaldo1998channel}. 

Therefore, in this article, a generalized Langevin equation is constructed to describe the process of population variation including the natural increase and migration; and it can be further transformed into the Fokker-Plank (F-P) equation to investigate the statistical characteristics of population migration. As is well known, the F-P equation is a drift-diffusion equation describing the diffusion of particles in an external potential field, and has been widely extended to investigate many other processes in which the movements of the objects demonstrate the similarity to the random diffusion (Markovian or non-Markovian) of a particle (see the typical representatives in Refs. \citen{chechkin2009fluctuation,chavanis2008nonlinear,alber2007continuous}). Here, the solution of the F-P equation presents an evolutionary picture of human societies in the aspects of population and GDP. The distribution of the regional populations originally obey the Gaussian-like distribution caused by the random migration. Subsequently, this distribution gradually transformed into a Shifted Power Law (SPL). Based on the Cobb-Douglas (C-D) production function, we suggest a relation function between population size and GDP, and derive the distribution of the regional GDPs. The distribution shows a reverse change from a SPL distribution into the Gaussian distribution that may be regarded as the prediction about the future of our society. If so, what does the Gaussian distribution of the regional GDPs imply? What causes the Gaussian distribution to occur? The following sections will present the detailed discussions. 

\section*{Results}

\subsection*{Statistical characteristics of the regional population sizes and GDPs}

\begin{figure*}
\epsfig{figure=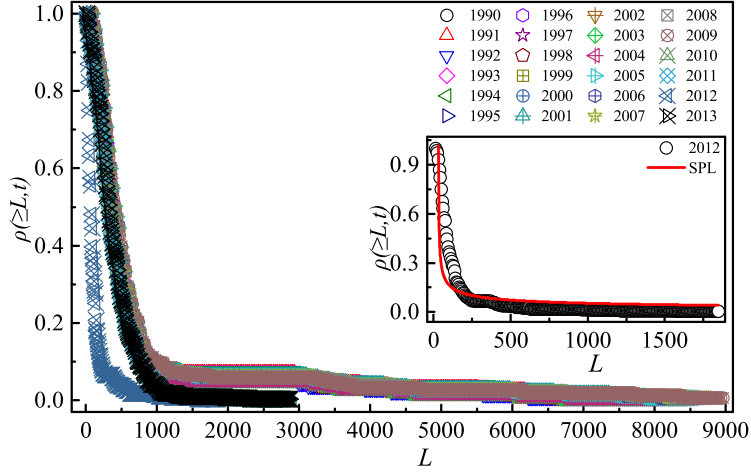,width=0.7\linewidth}
\caption{{\bf  Cumulative probability distribution of population sizes obtain from the China Statistical Yearbook from 1990 to 2013.} The inset presents the fitting by SPL relation for the distribution in 2012.}
\label{fig1}
\end{figure*}

Before presenting the theoretical discussions based on the similarity between human and natural substance, we perform a practical statistics to explore the characteristics of population distribution and the related economic regularity. We have collected some basic statistics including regional population size $L$ and GDP $Y$, of about 150 cities in China. We choose these cities for the relative complete data recorded in the China Statistical Yearbook from 1990 to 2013 (\url{http://www.stats.gov.cn}). Fig. \ref{fig1} presents the cumulative probability distribution of population sizes at each given time $t$(in one year) from 1990 to 2013, respectively. The inset re-displays the relation in 2012 chosen as one representative and fitted by the SPL

\begin{figure*}
\epsfig{figure=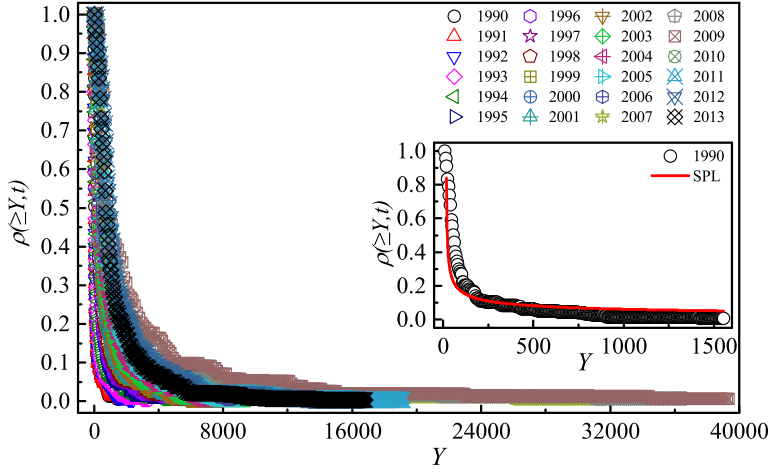,width=0.7\linewidth}
\caption{{\bf  Cumulative probability distribution of GDPs obtained from the China Statistical Yearbook from 1990 to 2013.} The inset presents the fitting by SPL relation for the distribution in 1990.}
\label{fig2}
\end{figure*}

\begin{eqnarray}\label{eq1}
\begin{aligned}
\rho(\geq L)\propto(L+a_{1})^{-b_{1}}.
\end{aligned}
\end{eqnarray}
Where $a_{1}=-31.673\pm0.344$, $b_{1}=0.425\pm0.018$. Fig. \ref{fig2} shows the cumulative probability distribution of GDPs in the same period. The representative relation in 1990 is also well fitted by the SPL and redisplayed by the inset, that is, the distribution function takes a similar form of Eq.\ref{eq1},

\begin{eqnarray}\label{eq2}
\begin{aligned}
\rho(\geq Y)\propto(Y+a_{2})^{-b_{2}}.
\end{aligned}
\end{eqnarray}
where $a_{2}=-18.141\pm0.296$, $b_{2}=0.406\pm0.0125$.

It is necessary to point out that these fitting results, SPL distributions, indicate that the evolution of the social system may be jointly driven by deterministic and stochastic dynamic factors according to the statement appeared in Refs.\citen{yan2013diversity} and \citen{simini2012universal}. This conclusion is also in accord with our previous discussions in natural system\cite{wang2012statistical} and social system\cite{Huo2016Statistical}. The understandings in our theoretical framework will present in the following sections.  

\subsection*{Stochastic Dynamical Equation for Evolution of Regional Population}
 
It is well known, the change of regional population is nothing more than the natural increase and migration, and the diffusion of population between regions owes much to the migration. The migration is often induced by higher income at the destination. Therefore, the regional population strongly depends on the regional economy. The economic factors, such as fixed capital investment, flowing capital investment, labor input, marginal propensity to save, potential capacity, etc., vary randomly. Therefore, the evolution of a regional population, $L$, is a typical stochastic process, and we suggest a generalized Langevin equation to describe this process. It reads in a general form,

\begin{eqnarray}\label{eq3}
\begin{aligned}
\displaystyle \frac{{\rm d}L}{{\rm d}t}=f(L,t)-g(L,t)\sigma(t)
\end{aligned}
\end{eqnarray}
where the first term on the right hand is deterministic corresponding to driving force, and the second stochastic corresponding to the Langevin force induced by the fluctuations of the related factors. Variable $\sigma(t)$ denotes the Gaussian white noise that is characterized by mean value $<\sigma(t)>=0$ and $\delta$-correlation $<\sigma(t)\sigma(t^{'})>=2D\delta(t-t^{'})$.

The apparent form of Eq. 1 can be obtained by correlating the well-known production function\cite{cobb1928theory} with the economic growth model\cite{Domar1945Capital,domar1947expansion}. The former stresses the dependence of the regional GDP, $Y$, on the determining factors such as labor input $L^{'}$, fixed capital investment $K$ and comprehensive technology factor $A$ (which is related to production efficiency), that is,

\begin{eqnarray}\label{eq4}
\begin{aligned}
Y=AL^{'\alpha}K^{\beta}.
\end{aligned}
\end{eqnarray} 
Where $\alpha$, $\beta$ is the elasticity coefficient for the labor output and the capital output, respectively. Here we suppose that the labor input is simply proportional to the population, so we have

\begin{eqnarray}\label{eq5}
\begin{aligned}
L^{'}=\eta L, 0<\eta<1.
\end{aligned}
\end{eqnarray}
The latter, the economic growth model, describes economic growth based on a pure quantitative relation linking output and input, which emphasizes that the time variation rate of GDP increases exponentially with time, and is related to the so-called marginal propensity to save   and potential social average productivity of investment $\xi$ in the following form

\begin{eqnarray}\label{eq6}
\begin{aligned}
\displaystyle \frac{{\rm d}Y}{{\rm d}t}=\frac{{\rm d}I}{{\rm d}t} \frac{1}{\lambda}=(\xi/\lambda)I_{0}e^{\xi     \lambda t}.
\end{aligned}
\end{eqnarray}
Where $I$, $I_{0}$ denotes the capital investment rate and its initial value, respectively. Obviously, the increment of GDP within time interval $t\rightarrow t+\Delta t$ is $A\eta^{-\alpha}(L+\Delta L)^{\alpha}K^{\beta}-A\eta^{-\alpha}L^{\alpha}K^{\beta}$ obtained from Eqs.\ref{eq3} and \ref{eq4}, and $(\xi/\lambda)I_{0}e^{\xi \lambda t}\Delta t$ obtained from Eq.\ref{eq6}. As a result, we have an equation

\begin{eqnarray}\label{eq7}
\begin{aligned}
(\xi/\lambda)I_{0}e^{\xi \lambda t}\Delta t=A\eta^{-\alpha}(L+\Delta L)^{\alpha}K^{\beta}-A\eta^{-\alpha}L^{\alpha}K^{\beta}.
\end{aligned}
\end{eqnarray}
As $\Delta L(\Delta t)$ has a very small value, the first term on the right hand of the equation can be expanded in Taylor series, and changed into the following form by remaining the linear term

\begin{eqnarray}\label{eq8}
\begin{aligned}
(\xi/\lambda)I_{0}e^{\xi \lambda t}\Delta t=A\alpha \eta^{-\alpha}L^{\alpha-1}K^{\beta} \Delta L.
\end{aligned}
\end{eqnarray}
As $\Delta t\rightarrow 0$ and therefore $\Delta L\rightarrow 0$, the equation can be transformed into the one

\begin{eqnarray}\label{eq9}
\begin{aligned}
\displaystyle \frac{{\rm d}L}{{\rm d}t}=\frac{\xi I_{0}e^{\xi \lambda t}\eta^{\alpha}}{\lambda \alpha AK^{\beta}}L^{\gamma}
\end{aligned}
\end{eqnarray}
by setting $\gamma=\alpha-1$. The right side of the equation is actually the deterministic term of Langevin equation, $f(L,t)$, in which each of the coefficients and variables can be taken as its mean value. Of course, each of the coefficients and variables fluctuates, so the instantaneous form of Eq.\ref{eq9} can be written as

\begin{eqnarray}\label{eq10}
\begin{aligned}
\displaystyle \frac{{\rm d}L}{{\rm d}t}=\frac{(\xi \pm \Delta \xi)(I_{0} \pm \Delta I_{0})(\eta \pm \Delta \eta)^{(\alpha \pm \Delta \alpha)}e^{(\xi \pm \Delta \xi)(\lambda \pm \Delta \lambda)t}}{(\lambda \pm \Delta \lambda)(\alpha \pm \Delta \alpha)(A \pm \Delta A)(K \pm \Delta K)^{(\beta \pm \Delta \beta)}}L^{\gamma},
\end{aligned}
\end{eqnarray}
where $\Delta \xi$, $\Delta I_{0}$, $\Delta \eta$, $\Delta \alpha$, $\Delta \lambda$, $\Delta A$, $\Delta K$, $\Delta \beta$ denotes the amplitude of the fluctuation of $\xi$, $I_{0}$, $\eta$, $\alpha$, $\lambda$, $A$, $K$, $\beta$, respectively. Truncating appropriately the Taylor series expansion for each term, the equation changes to the following one,

\begin{eqnarray}\label{eq11}
\begin{aligned}
\displaystyle \frac{{\rm d}L}{{\rm d}t}=\frac{\xi I_{0}\eta^{\alpha}e^{\xi \lambda t}}{\lambda \alpha AK^{\beta}}L^{\gamma}(1 \pm (\frac{\Delta \xi}{\xi}+\frac{I_{0}}{I_{0}}+\alpha \frac{\Delta \eta}{\eta}+\Delta \alpha+\Delta \xi +\Delta \lambda-\frac{\Delta \lambda}{\lambda}-\frac{\Delta \alpha}{\alpha}-\frac{\Delta A}{A}-\beta \frac{\Delta K}{K}-\Delta \beta)).
\end{aligned}
\end{eqnarray}
By setting $\phi=\displaystyle \frac{\xi I_{0}\eta^{\alpha}e^{\xi \lambda t}}{\lambda \alpha AK^{\beta}}$ and $\varepsilon=\displaystyle \frac{\Delta \xi}{\xi}+\frac{I_{0}}{I_{0}}+\alpha \frac{\Delta \eta}{\eta}+\Delta \alpha+\Delta \xi +\Delta \lambda-\frac{\Delta \lambda}{\lambda}-\frac{\Delta \alpha}{\alpha}-\frac{\Delta A}{A}-\beta \frac{\Delta K}{K}-\Delta \beta)$, the equation is rewritten as 

\begin{eqnarray}\label{eq12}
\begin{aligned}
\displaystyle \frac{{\rm d}L}{{\rm d}t}=\phi L^{\gamma} \pm \phi \varepsilon L^{\gamma},
\end{aligned}
\end{eqnarray}
here $\pm \varepsilon$ denotes pure noise caused by the relative fluctuations, and the noise is in terms of $\pm \phi \varepsilon$. Comparing Eq.\ref{eq3} with Eq.\ref{eq12}, one can easily obtain the two terms, 

\begin{eqnarray}\label{eq13}
\begin{aligned}
f(L,t)=\phi L^{\gamma}, g(L,t)=L^{\gamma}.
\end{aligned}
\end{eqnarray}

Thus, a generalized Langevin equation describing the random growth of regional population is established. It is very clear that such a stochastic equation cannot predict the migration precisely, but can only describe the migration by probability. Therefore, a transform, from the generalized Langevin equation to the F-P equation, is necessary to carry out.  

\subsection*{Distribution of Regional Population Sizes}

The transform can be conducted as follows. Firstly, one can obtain $L(t+\tau)-L(t)$ via integrating Eq.\ref{eq3} within a small time interval $\tau$ (in $t$-$t+\tau$) along the evolutionary path or integral path. Secondly, one can get n-order moment, for any possible n, as functions $f(L,t)$ and $g(L,t)$ in the expression of $L(t+\tau)-L(t)$ are expanded in series in the neighborhood of $L(t)$, that is,

\begin{eqnarray}\label{eq14}
\begin{cases}
\langle L(t+\tau)-L(t)\rangle =\lbrace f(L,t)+Dg^{\prime}(L,t)g(L,t)\rbrace \tau +O(\tau^{2}),\\
\langle [L(t+\tau)-L(t)]^{2}\rangle =2Dg^{2}(L,t)\tau +O(\tau^{2}),\\
\langle [L(t+\tau)-L(t)]^{n}\rangle =0 \qquad n\geqslant 3.
\end{cases}
\end{eqnarray}
Where $g^{\prime}(L,t)$ is the derivative of $g(L,t)$ with respect to $L$. Finally, the F-P equation can be derived from the moments above via substituting them into the so-called Kramers-Moyal expansion holding over the linear terms. And then we have equation ,

\begin{eqnarray}\label{eq15}
\begin{aligned}
\frac{\partial}{\partial t}\rho(L_{0}|L;t)=-\frac{\partial}{\partial L}\lbrace f(L,t)+Dg^{\prime}(L,t)g(L,t)\rbrace \rho(L_{0}|L;t)+D\frac{\partial^{2}}{\partial L^{2}}\lbrace g^{2}(L,t)\rho(L_{0}|L;t)\rbrace 
\end{aligned}
\end{eqnarray}
to describe the time evolution of the probability distribution of population sizes. Where $\rho(L_{0}|L;t)$ denotes the transition probability to measure the probability of a region with population $L_{0}$ at $t=0$ becoming the one with population $L$ after a while $t$.

Now let’s find a general solution of Eq.\ref{eq15}, $\rho(L^{\prime},t|L,t+\tau)$, which denotes the transition probability of the system from time $t$ with population $L^{\prime}$ to time $t+\tau$ with population $L$ in a small time interval $\tau$. It can be obtained by integrating over $\tau$, 
 
\begin{eqnarray}\label{eq16}
\begin{aligned}
\rho(L^{\prime},t|L,t+\tau)=[1-\frac{\partial}{\partial L}(\varphi L^{\gamma}+\gamma D L^{2\gamma-1})\tau +D\frac{\partial^{2}}{\partial L^{2}}L^{2\gamma}\tau +O(\tau^{2})]\delta(L-L^{\prime}).
\end{aligned}
\end{eqnarray}
The terms of order being equal and higher than 2 may be ignored due to the very small values. So, Eq.\ref{eq16} is simplified as

\begin{eqnarray}\label{eq17}
\begin{aligned} 
\rho(L^{\prime},t|L,t+\tau)\approx [1-\frac{\partial}{\partial L}(\varphi L^{\gamma}+\gamma D p^{2\gamma-1})\tau +D\frac{\partial^{2}}{\partial L^{2}}L^{2\gamma}\tau]\delta(L-L^{\prime}).
\end{aligned}
\end{eqnarray}
One may note that the polynomial in the bracket can be regarded as the linear function of time interval $\tau$. It can also be treated as a Taylor series expansion of an exponential function, of which the terms of order 0 and 1 are retained. Therefore, the transition probability can be approximatively "returned" the corresponding exponential function, that is,

\begin{eqnarray}\label{eq18}
\begin{aligned} 
\rho(L^{\prime},t|L,t+\tau)\approx exp[-\frac{\partial}{\partial L}(\varphi L^{\gamma}+\gamma D L^{2\gamma-1})\tau +D\frac{\partial^{2}}{\partial L^{2}}L^{2\gamma}\tau]\delta(L-L^{\prime}).
\end{aligned}
\end{eqnarray}
Due to the fact that the probability described by the exponential function is not normalized, the approximately equal sign in the above formula can be changed into an equal sign. After that, introducing the Fourier integral to express the $\delta$-function, one obtains 

\begin{eqnarray}\label{eq19}
\begin{aligned} 
\rho(L^{\prime},t|L,t+\tau)=exp[-\frac{\partial}{\partial L}(\varphi L^{\gamma}+\gamma D L^{2\gamma-1})\tau +D\frac{\partial^{2}}{\partial L^{2}}L^{2\gamma}\tau]\frac{1}{2\pi}\int_{-\infty}^{\infty}e^{-iu(L-L^{\prime}){\rm d}u}.
\end{aligned}
\end{eqnarray}
According to the differential relation of the Fourier transform, $\frac{d^{n}F(\omega)}{d \omega^{n}}=(-1)^{n}[t^{n}f(t)]$, the equation can be written in the form 

\begin{eqnarray}\label{eq20}
\begin{aligned}
\rho(L^{\prime},t|L,t+\tau |)=&\frac{1}{2\pi}\int_{-\infty}^{\infty}exp[-iu(\varphi L^{\gamma}+\gamma D L^{2\gamma -1})\tau -u^{2}D L^{2\gamma}\tau+iu(L-L^{\prime})]{\rm d}u\\
=&\frac{1}{2\pi}\int_{-\infty}^{\infty}exp[-D L^{2\gamma}\tau(u^{2}-iu\frac{[(L-L^{\prime})-(\varphi L^{\gamma}+\gamma D L^{2\gamma -1})\tau]}{D L^{2\gamma}\tau}\\
&+\frac{[(L-L^{\prime})-(\varphi L^{\gamma}+\gamma D L^{2\gamma -1})\tau]^{2}}{4(D L^{2\gamma}\tau)^{2}})-\frac{[(L-L^{\prime})-(\varphi L^{\gamma}+\gamma D L^{2\gamma -1})\tau]^{2}}{4 D L^{2\gamma}\tau}]{\rm d}u\\
=&\frac{1}{2\pi}\int_{-\infty}^{\infty}exp[-D L^{2\gamma}\tau(u-i\frac{[(L-L^{\prime})-(\varphi L^{\gamma}+\gamma D L^{2\gamma -1})\tau]}{2 D L^{2\gamma}\tau})^{2}-\frac{[(L-L^{\prime})-(\varphi L^{\gamma}+\gamma D L^{2\gamma -1})\tau]^{2}}{4 D L^{2\gamma}\tau}]{\rm d}u.
\end{aligned}
\end{eqnarray}
Using the Gaussian integral, $\int_{-\infty}^{\infty}e^{-x^{2}}dx=\sqrt{\pi}$, one therefore has

\begin{eqnarray}\label{eq21}
\begin{aligned} 
\rho(L^{\prime},t|L,t+\tau)=&\frac{1}{2\sqrt{\pi D L^{2\gamma}\tau}}exp(-\frac{[(L-L^{\prime})-(\varphi L^{\gamma}+\gamma D L^{2\gamma -1})\tau]^{2}}{4 D L^{2\gamma}\tau})\\
=&\frac{L^{-\gamma}}{2\sqrt{\pi D \tau}}exp(-\frac{[(L-L^{\prime})-(\varphi L^{\gamma}+\gamma D L^{2\gamma -1})\tau]^{2}}{4 D L^{2\gamma}\tau}).
\end{aligned}
\end{eqnarray}

By setting $L^{\prime}=0$, $t=0$ (the starting point of the evolution is assumed to extrapolate backward to the "primitive society"), and $\tau\rightarrow t$(extending the small time situation to an arbitrary time situation), so transition probability $\rho(L^{\prime},t|L,t+\tau)$ can be simplified as $\rho(L,t)$, and then its expression is changed to 

\begin{eqnarray}\label{eq22}
\begin{aligned}
\rho(L,t)=\frac{L^{-\gamma}}{2\sqrt{\pi D t}}exp(-\varphi \gamma L^{\gamma -1}t+D\gamma (2\gamma -1)L^{2\gamma -2}t-\frac{[L^{1-\gamma}-\varphi t+\gamma D L^{\gamma -1}t]^{2}}{4 D t}).
\end{aligned}
\end{eqnarray}

In the following discussions, the values of main parameters are set as $\varphi=0.285$, $\gamma=1.47$, $D=0.0005$. Obviously, this solution is characterized by a product of two functions. One is in power form and the other in exponent form. This distribution is called Generalized Exponential Power Distribution (GEPD)\cite{A2002Evolution,Chang2007,Raup1986Biological,Laherr1998Stretched,wang2012statistical,Huo2016Statistical}. Similar distribution characteristics can be observed in many real systems, such as temporal and spatial activities of human, biological system, river networks and so on\cite{Laherr1998Stretched}. As discussed in Ref. \citen{wang2012statistical}, such a distribution law has two equivalent mathematical formulations. One is known as the SPL, expressed as $\rho(x)\propto(x+a)^{-b}$, the other is described as the TPL that is manifestation of $\rho(x)=\mu(x^{\mu-1}/x^{\mu}_{0})exp(-(x/x_{0})^{\mu})$. The characteristic parameter for the former and the latter is $a$ and $\mu$, respectively. When $a=0$($\mu \rightarrow 0$), the SPL(TPL) relation tends to be a typical power-law relation, whereas when $a\rightarrow \infty$($\mu=1$) the SPL(TPL) relation reduces to a typical exponential distribution. 

\begin{figure*}
\epsfig{figure=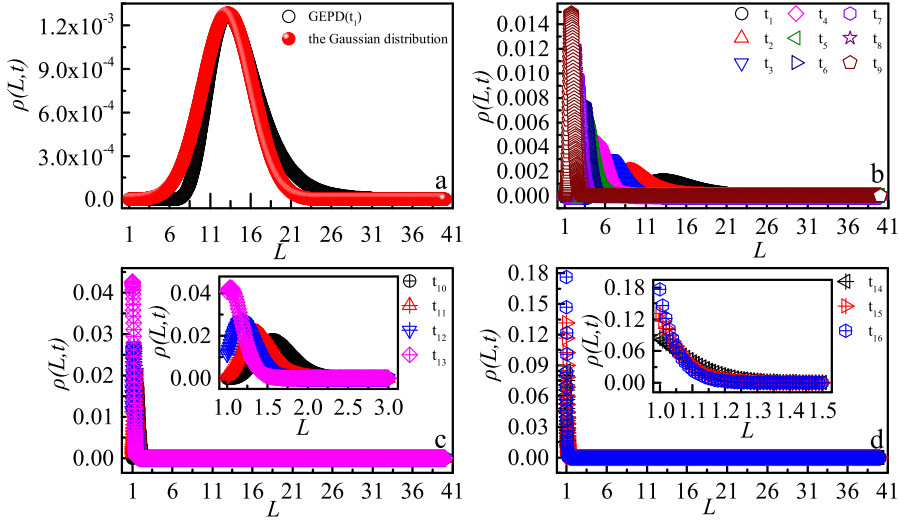,width=0.7\linewidth}
\caption{{\bf Distribution of the regional populations varies with time.} (a) presents the distribution at $t=1$, the distribution is close to the normal Gaussian distribution; (b) presents the distribution from $t=1$ to $t=9$; (c) presents the distribution from $t=10$ to $t=13$, its enlargement is shown by the inset; (d) presents the distribution from $t=14$ to $t=16$, their enlargement is also shown by the inset.}
\label{fig3}
\end{figure*}

\begin{figure*}
\epsfig{figure=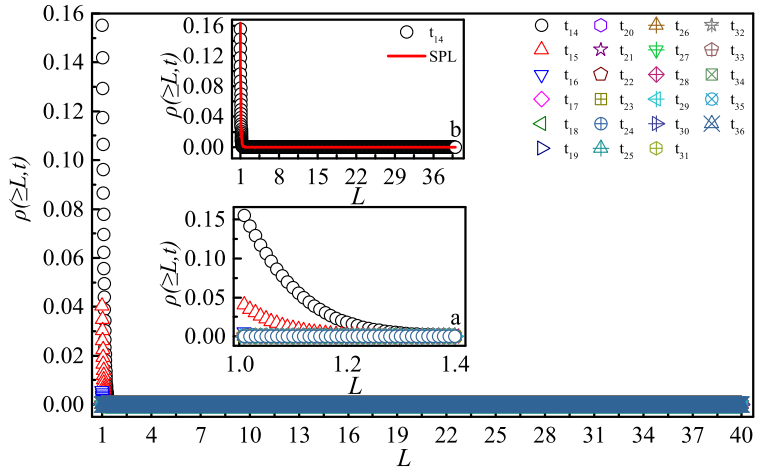,width=0.7\linewidth}
\caption{{\bf Cumulative probability distribution of populations.} The inset (a) is the partial enlargement and the inset (b) the fitting of that at $t=14$, obeying the SPL.}
\label{fig4}
\end{figure*}

Here Fig. \ref{fig3} depicts the final analytical result presented by Eq. \ref{eq22}, the probability distribution of regional population sizes varies with time. This result may reproduce some historical profiles of the human society, that is, exhibit the regional population evolution in the process of the natural increase and migration, as well as the related economy development. Fig. \ref{fig3}(a) shows a typical random distribution similar to the Gaussian distribution (the comparison between the distribution and the standard Gaussian distribution is shown in the Figure), which indicates that human distributes evenly and the "society" runs without structure in the early history. This may be attributed to the low natural population increase rate and the rare and blind migration. Fig. \ref{fig3}(b) demonstrates a gradual change of the distribution with time. The distribution curve loses its early symmetry with respect to the peak point. And the peak point moves to the left. Fig. \ref{fig3}(c) shows the distribution reaches a critical point at $t=13$. The supercritical distribution completely transforms into the one monotonously descending with population size $L$. Fig. \ref{fig3}(d) shows the details. Then a question may naturally occur, i.e., what kind is this monotonously descending distribution? Before answering the question, we should check the validation of our theoretical framework via comparing the theoretical result with that obtained practically. So, one has to perform an integration for Eq.\ref{eq22} from $L$ to infinity to yield the cumulative probability distribution of population sizes

\begin{eqnarray}\label{eq23}
\begin{aligned}
\rho(\geq L,t)=\int_{L}^{\infty}\rho(L,t){\rm d}L.
\end{aligned}
\end{eqnarray}
Fig. \ref{fig4} presents the cumulative probability in the supercritical situation at $t=14$. The fitting red solid curve suggests a perfect SPL

\begin{eqnarray}\label{eq24}
\begin{aligned}
\rho(\geq L,t)\propto(x+a_{3})^{-b_{3}}
\end{aligned}
\end{eqnarray}
with $a_{3}=0.124\pm0.00039$, $b_{3}=6.31\pm0.0091$. It is clear that the theoretical result is in qualitative accordance with the observed result and that the answer is revealed. It also implies that our theoretical framework can describe the characteristics of such spatial behavior of human. It seems now we can sketch out a dynamical profile of human population distribution evolving with time. In the early period of the evolutionary process, supposing it is the "primitive society", the population migrates or proliferates in a disorderly manner, and therefore leads to the Gaussian distribution. As time goes on, the cluster effect in some regions begins to emerge, so the even "society" signalled by the Gaussian distribution is destroyed, and is gradually transforming to the SPL distribution. This also marks that the society is evolving from disorder to order. Finally, the distribution turns into the SPL, which means that the society has evolve into an order state with a structure of highly inhomogeneous regional populations. In other words, the SPL dominating the population distribution is a sign of a modern society with a developed structure. 

\subsection*{Distribution of Regional GDPs}

\begin{figure*}
\epsfig{figure=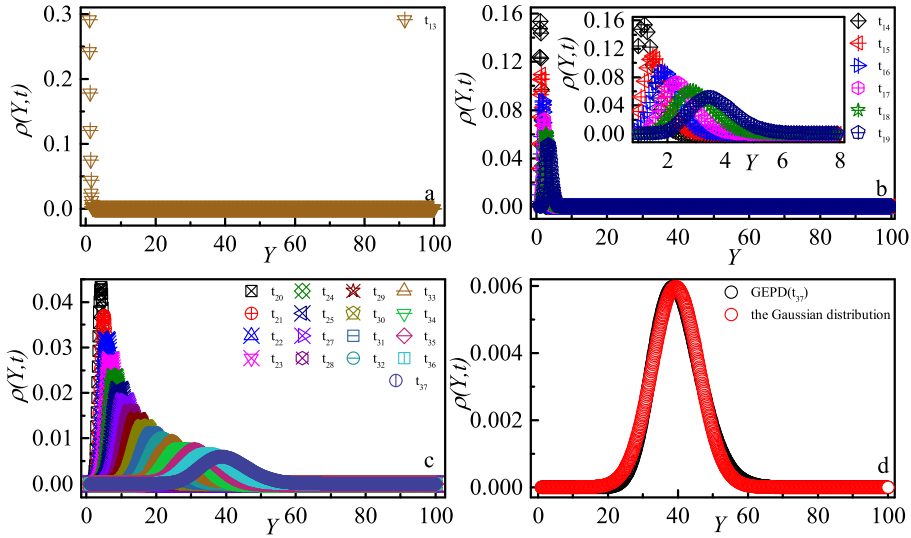,width=0.7\linewidth}
\caption{{\bf Distribution of the regional GDPs varies with time.} (a) presents the distribution at critical point $t=13$, the distribution is close to the SPL; (b) presents the distribution from $t=14$ to $t=19$, the distribution turns into the normal Gaussian distribution at $t=19$; (c) presents the distribution from $t=20$ to $t=37$; (d) re-presents the distribution at $t=37$,the comparison of it with the normal Gaussian distribution indicates that the distribution keeps well in the Gaussian fashion.}
\label{fig5}
\end{figure*}

It is generally accepted that the evolution of the regional populations is mainly driven by the benefit pursuing of human to improve the living conditions. So the population closely relates to the economic development level represented by GDP. It especially is so for the modern society. The widely approved relation is known as the C-D production function that can be approximately written as $Y\propto L^{\alpha}$  . One can suggest a relation function for the population and GDP using a $\delta$-function, 

\begin{eqnarray}\label{eq25}
\begin{aligned}
r(L,Y)=\delta(L-Y^{-\alpha}).
\end{aligned}
\end{eqnarray}
With which one can obtain the distribution of the regional GDPs via performing the following integration 

\begin{eqnarray}\label{eq26}
\begin{aligned}
\rho(Y,t)=&\int \rho(L,t)r(L,Y){\rm d}L\\
=&\frac{Y^{\alpha \gamma}}{2\sqrt{\pi D t}}exp(-\varphi \gamma Y^{-\alpha(\gamma -1)}t+D\gamma (2\gamma -1)Y^{-\alpha(2\gamma -2)}t-\frac{[Y^{-\alpha(1-\gamma)}-\varphi t+\gamma D Y^{-\alpha(\gamma -1)}t]^{2}}{4 D t}).
\end{aligned}
\end{eqnarray}
It is necessary to point out that the relation function between the regional population and GDP relates only to those in supercritical situation. Equally, the integral, distribution of the regional GDPs, is applied only to the supercritical situation. One may say that if the distribution of the regional GDPs can describe the current state, it also can predict the future of the world. The calculating result of Eq. \ref{eq25} is shown by Fig. \ref{fig5} as $\alpha=1.2$ and the other parameters are the same as those in calculating the distribution of population sizes. Fig.\ref{fig5} (a) shows the distribution of the regional GDPs at the critical point, $t=13$, which can be perfectly fitted by a SPL. However, it is amazing that the SPL distribution here does not like the distribution of population, staying the way. As is shown by Figs. \ref{fig5}(b) and (c), the distribution is destroyed, and a peak appears on the curve as it is in the supercritical situation. As time goes on, the distribution turns into the Gaussian distribution, Fig. \ref{fig5}(d) proves that the distribution of the regional GDPs is a perfect Gaussian distribution. Furthermore, the peak of the Gaussian distribution moves to the right, and drops and broadens meanwhile. The movement to the right implies the increase of the GDP as a whole, the dropping and broadening variation of the peak indicates the homogenization of the distribution. This varying trend may accord with the fact that almost everyone and every region is inclined to seek wealth in the developing process of human society. Perhaps, the Gaussian distribution may be the economic status of our world in future.

  It may be important if one can know the cause why the change from the SPL distribution to the Gaussian distribution can occur. However, it is very difficult to provide a physical answer to this question due to the lack of dynamical detail. We can only understand the change mathematically.

The distribution, Eq. \ref{eq25}, includes three factors, $Y^{\alpha\gamma}$, $\frac{1}{2\sqrt{\pi D t}}exp(-\frac{[Y^{-\alpha(\gamma -1)}t-\varphi t+\gamma D Y^{-\alpha(\gamma -1)t}]^{2}}{4 D t}$ and $exp(-\varphi \gamma Y^{-\alpha(\gamma -1)}t+D\gamma (2\gamma -1)Y^{-\alpha(2\gamma -2)}t)$. The first factor indicates a growth in power-law. The second factor is the leading factor and actually the Gaussian distribution, which is modulated by the first and the third ones. The second and the third are similar to those corresponding factors in Eq. \ref{eq22}. However, the first varies in the direction opposite to that of Eq.  \ref{eq22}, decaying in power-law. This can mathematically interpret the cause why the GDP distribution varies in the direction (from the SPL to the Gaussian in the supercritical situation) of which is completely opposite to that of the population distribution (from the Gaussian to the SPL in the subcritical situation). 

\section*{Discussion}

In this paper, the empirical investigation on the regional populations and GDPs is carried out, and reveals that the distribution of the regional populations and that of the regional GDPs obey the similar regularities, the SPL. 

To understand these distributions theoretically, we suggest a generalized Langevin equation to describe the evolution of regional population driven by the deterministic economic factors such as fixed capital investment, flowing capital investment, marginal propensity to save, potential capacity, as well as their random fluctuations. The generalized Langevin equation is transformed into the F-P equation, by a mathematical technique, to describe the evolution of the population distribution. The solution shows a gradual transformation of the distribution from the normal Gaussian distribution at the evolutionary beginning to the SPL one as it is beyond a fixed time $t=13$. The fixed time is actually a critical point of time. As it is supercritical, the SPL sets up and then remains qualitatively unchanged. Obviously, this situation corresponds to the current practical observations. 

Relating the regional GDP to the regional population by introducing a relation function based on the well-known C-D production function, we obtain the distribution of the regional GDPs. The distribution demonstrates the variation, in the supercritical situation, from the SPL to the Gaussian. 

As far as the meanings of these distributions are concerned, the variation of the population distribution implies that the world evolves from a disorder or an unstructured "primitive society" marked by the Gaussian distribution to an order and highly inhomogeneous structured "modern society" characterized by the SPL distribution. How can this SPL distribution set up? The answer may lie in the development pattern of human society, attracting the population by economy/resources and developing the economy/richening the resources by population reversely. Therefore, the regions of rich resources become the population gathered regions and the social structure expressed by the SPL distribution of populations gradually sets up. The GDP distribution indicates a transformation in the supercritical situation, from the SPL to the Gaussian. This implies that our society is currently in the time point near to the critical point due to the two distributions of populations and GDPs are all the SPL and qualitatively in accordance with the observations. And then one can predict that our society might evolve into the one, in future, dominated by the Gaussian distribution of regional GDPs. Furthermore, the evolution of the world might tend to becoming "common prosperity" denoted by the distribution peak moving to the right, the peak value dropping and the range broadening. Perhaps, someone raises the question of whether there is a contradiction between the two distributions in the supercritical situation--one is the SPL, the other is the Gaussian. The answer may be no due to the basic fact that the growth mode of the GDP is gradually changes from the previous resource-based to the future science-technology-based. With scientific and technological advancements, the close correlation between the production and population becomes invalid; the regional difference in the economy scale reduces or eliminates. Therefore, the relative uniform distribution of the regional GDPs denoted by the Gaussian distribution will occur naturally. Of course, this change might not change the distribution of the regional populations, maintaining the SPL.

Merging the evolution of the population distribution with that of the GDP distribution, a historical picture of human society spreads in front of us. This might have important academic values for enhancing the knowledge about the evolution of our society, population and the related economy. 

\section*{Acknowledgements}

This study is supported by National Natural Science Foundation of China (Grant Nos. 11265011 and 11665018).

\section*{Author contributions statement}

All authors participated in the design of the project. J. H. and X. -M. W. wrote the manuscript. J. H. and P. W. collected data and analysis. J. H., X. -M. W. and R. H. discussed the model. J. H. and X. -M. W. did theoretical analysis. X. -M. W. supervised the research.

\end{document}